# *Neural network design of multilayer metamaterial for temporal differentiation*

*Tony Knightley[1,2], Alex Yakovlev[2] and Victor Pacheco-Peña[1*]*

[1] School of Mathematics, Statistics and Physics, Newcastle University, Newcastle Upon Tyne, NE1 7RU, United Kingdom
[2] School of Engineering, Newcastle University, Newcastle Upon Tyne, NE1 7RU, United Kingdom
*email: victor.pacheco-pena@newcsatle.ac.uk

**Controlling wave-matter interactions with metamaterials (MTMs) for the calculation of mathematical operations has become an important paradigm for analogue computing given their ability to dramatically increase computational processing speeds. Here, motivated by the importance of performing mathematical operations on temporal signals, we propose, design and study multilayer MTMs with the ability to calculate the derivative of incident modulated temporal signals, as an example of a significant computing process for signal processing. To do this, we make use of a neural network (NN) based algorithm to design the multilayer structures (alternating layers of indium tin oxide (ITO) and titanium dioxide ($TiO_2$)) that can calculate the first temporal derivative of the envelope of an impinging electromagnetic signal at telecom wavelengths (modulated wavelength of 1550 nm). Different designs are presented using multiple incident temporal signals including a modulated Gaussian as well as modulated arbitrary functions, demonstrating an excellent agreement between the predicted results (NN results) and the theoretical (ideal) values. It is shown how, for all the designs, the proposed NN-based algorithm can complete its search of design space for the layer thicknesses of the multilayer MTM after just a few seconds, with a low mean square error in the order of (or below) $10^{-4}$ when comparing the predicted results with the theoretical spectrum of the ideal temporal derivative.**

# Introduction

Owing largely to the rapid development of semiconductor-based technologies throughout the 20$^{th}$ century, digital electronics is easily the dominant paradigm in computing today. These ubiquitous systems are well-known for their compact size and strong potential for re-programmability, making them ideal for general computational tasks [1]. However, physical limitations (such as quantum tunnelling) imposed on the size and separation of individual transistors in semiconductor-based circuits means that performance gains may soon reach their maximum potential [2]. Additionally, parasitic capacitances appearing in these devices can lead to challenges in energy consumption and speed during the charging/discharging processes involved when using them as switching devices for computing [3]. While it is expected that this technology remains with us for many years to come and research is underway to overcome challenges and further improve its performance, ever-increasing demand for faster and more energy efficient computations therefore necessitates the introduction of alternative computing paradigms. Some notable examples that have been explored so far include quantum computing [4], TEM pulse switching [5]–[8] and computing with solitons [9], to name a few. In this realm, analogue wave-based computing, as an alternative to current electronic-based computing systems, offers inherently parallel processing capabilities for fast, energy efficient computation on specialised tasks such as high throughput image processing [10], [11], equation solving [12]–[14] and machine learning [15].

Another prominent area of research where this paradigm shift is taking off is in the field of metamaterials (MTMs) (and metasurfaces, MTSs, as their 2D counterparts) [16], [17]. MTMs and MTSs are artificially engineered structures that enable enhanced and tailored wave-matter interactions both in space and time [18]–[22] with responses not easily available in nature such as epsilon near-zero or negative refractive index [17], [23]–[25]. For the past two decades, the scientific community has reported many interesting applications using MTMs across different branches of science and engineering. For example, MTM-based antennas for high gain 5G communications [26], miniaturised optical neural networks for real-time object recognition [27], MTS-based sensors at terahertz frequencies [28] and invisibility cloaking devices [29], to name a few. Recently, the concept of wave-based analogue computing with MTMs has also been introduced [30]. In this emerging branch of MTM research, the interactions between an incoming wave and the multiple meta-atoms (scatterers) can be exploited to perform some mathematical operation of choice (e.g. differentiation, integration, convolution, etc.) with several ground-breaking designs being recently proposed and experimentally demonstrated [30]–[32]. For example, the MTS approach in [30] which can be thought of as the MTS analogue to the 4F correlator (more commonly associated with Fourier optics). These devices have been demonstrated to calculate a desired mathematical operation in the spatial frequency domain. Another prominent approach is the Green's function method which involves engineering the structural parameters of a multilayer MTM such that the transfer function approximates the Green's function of a desired mathematical operation [30], [33]–[37]. In addition to these examples, other interesting design approaches recently reported include MTS diffraction gratings [38]–[40] and computational plasmonic systems [41]–[43], among others.

To date, implementations of MTM-based analogue computing has mostly been focused on performing mathematical operations in the spatial domain using spatially modulated monochromatic electromagnetic (EM) waves [30], [38]–[40], [44]–[46]. Acoustic MTM-based analogue computing has also been investigated for both spatial [47] and temporal [48], [49] domains with other works exploiting spatiotemporal concepts for computing [50], [51]. However, there has so far been limited research focusing on specialised MTM-based optical analogue computing devices for temporal signals. This could be attributed to additional challenges introduced by the real dispersive nature of materials within the frequency range contained in the incident temporal signal. At microwave frequencies, analogue differentiator of temporal signals have been demonstrated with some remarkable examples using programable MTM-based structures [52].

While the design process of MTMs and MTSs can involve solving computationally intensive numerical/analytical calculations, machine learning (ML) techniques have recently been gaining a lot of popularity as an alternative approach [53]. Once purely a branch of computer science, ML has had much success in applications such as computer vision and natural language processing [54]. More recently, ML methods – and in particular, neural networks (NNs) – have gained popularity in many other branches of science and engineering as a novel data-driven research method. Examples of this include predicting 3D protein structures [55], quantum molecular wavefunctions [56] as well as cancer detection and prognosis [57]. Additionally, ML methods have been applied in the field of MTMs to expedite the design process and obtain solutions that may be difficult to achieve using traditional approaches [58]. In particular, NNs have had remarkable success with implementing the inverse design of MTMs [59]–[63]. However, inverse design techniques may present an important challenge in terms of the required spectral response (transmission or reflection, for instance) of the designed MTM: for a MTM structure with many potential parameterisations, it is possible that multiple candidates correspond to the same (or at least similar) optical spectral responses [59]. Therefore, in general, degeneracy is a key challenge for NN-enabled inverse design of MTMs. Some more advanced NN architectures have been utilised to address this degeneracy including the tandem network [64] and generative adversarial networks (GANs) [65], [66]. In addition to inverse design, it has also been shown that NNs can be trained for accelerated forward predictions [67]; i.e. to predict the optical response of a predefined MTM structure. Furthermore, as it will be demonstrated below, NN-accelerated forward predictions of optical responses can be combined with an iterative algorithm for efficiently obtaining an optimised solution within a specified design space. Importantly, this approach does not suffer from the degeneracy problem of inverse design methods as any given MTM design corresponds to just one optical response [59].

Inspired by the potential of MTM-based optical analogue computing for performing fast and efficient calculations on temporal signals, here we leverage ML methods to design and study a multilayer MTM that has the ability to perform temporal differentiation of a modulated EM Gaussian pulse via alternating Indium tin oxide (ITO) and Titanium dioxide ($TiO_2$) layers. This is carried out within the spectral wavelength range of 1428 – 1695 nm (177–210 THz) with a modulated wavelength of 1550 nm (193.54 THz), i.e., telecommunication wavelengths. Here, we make use of a fully connected NN for expedited forward predictions together with an iterative algorithm to find optimised layer thicknesses. Additionally, as proof of generalisability, we use our trained NN to identify a second design and apply this to obtain the derivative of two arbitrary temporal signals. In all three cases, we obtained optimised designs with a mean squared error (MSE) of the order $10^{-4}$ or below. These results may enable fast processing of mathematical operations and equation solving using temporal signals.

## Design Process

### Materials and design parameters

As mentioned in the introduction, our goal is to engineer a multilayer MTM using NN methodology such that the reflected signal represents the temporal derivative of an incoming modulated Gaussian pulse under normal incidence (see a schematic representation in Figure 1a and the steps to calculate the derivative in Figure 1c). To enable a compact design of our multilayer MTM, we choose a total of 13 alternating layers of Indium Tin Oxide (ITO) and Titanium Dioxide ($TiO_2$) starting and ending with ITO (see Figure 1a and Figure 2a). The multilayer MTM is designed to work at telecommunication wavelengths within the spectral range of ~1428 nm and ~1695 nm (i.e., 177 THz – 210 THz) with a central wavelength of 1550 nm (193.54 THz). Within this frequency range, the permittivity of $TiO_2$ varies from 6.004 to 6.037. Hence, from now on and for the sake of simplicity, we consider a nondispersive permittivity for $TiO_2$ with a value of $\varepsilon_{TiO_2} = 6.02$. For ITO, a full dispersive model is implemented (see Figure 1b) with,

$$\varepsilon_{ITO} = \varepsilon_{inf} - \frac{\omega_p^2}{\omega^2 + i\omega\Gamma_d} \quad (1)$$

where $\varepsilon_{inf} = 3.91$, $\omega_p = 2.65\times10^{15}$ rad s$^{-1}$, $\Gamma_d = 2.05\times10^{14}$ rad s$^{-1}$ and $\omega$ the operational frequency in rad s$^{-1}$ [68]. The ITO layers are constrained to be between 40 and 70 nm while the TiO$_2$ range from 200 to 300 nm. These parameters were chosen considering preliminary results indicating a good likelihood of finding an appropriate design for obtaining the correct spectrum of the temporal derivative of an incident signal (with a modulated frequency at the telecommunications wavelength of 1550 nm (not shown here)). All layers are presumed infinite in the transverse directions. With this configuration, and as it will be shown below, the reflection spectrum of an incident signal can be manipulated by varying the layer thicknesses (dimension along the propagation axis) within their respective ranges.

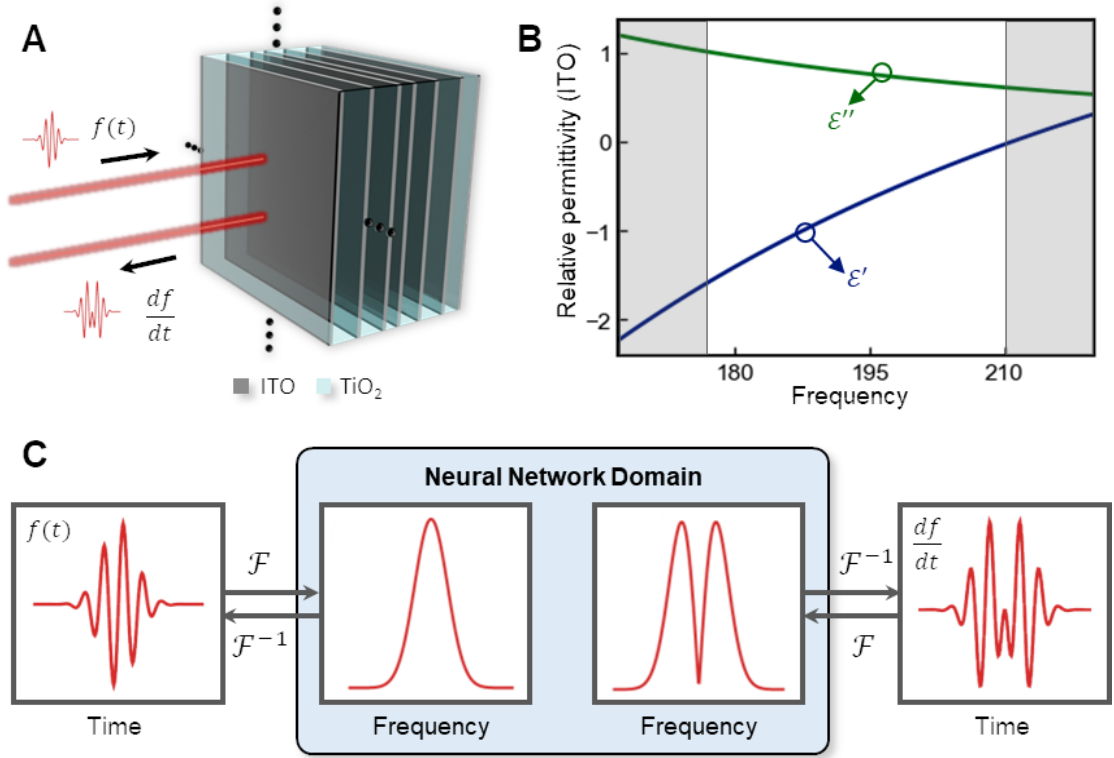

**Fig. 1 | Multilayer metamaterial (MTM) that solves for the derivative of a temporal Gaussian function.** (a) a schematic diagram of the proposed MTM to be designed below. (b) the complex relative permittivity of ITO (ε' real, and ε'' imaginary components) within the spectral range of interest (177 - 210 THz). (c) from left to right, shows the process of transforming the incident temporal signal into the frequency domain within which our NN predicts an appropriate MTM design, then back to the time domain. $\mathcal{F}$ and $\mathcal{F}^{-1}$ refer to the Fourier and inverse Fourier transforms, respectively.

*Fully connected NN design and training*

In order to obtain an appropriate set of layer thicknesses, we first trained a fully connected NN to predict the reflection spectra for MTM designs conforming to the specifications in the previous section (see Figure 2b for a diagram of the NN architecture). We used 13 nodes in the input layer corresponding to the ITO and TiO$_2$ thicknesses (blue $x_i$ nodes) and 200 output nodes representing data points in the predicted spectra (red $y_j$ nodes). It was found that one hidden layer with 128 nodes was sufficient for making accurate predictions. To train the NN, we generated a dataset of 50000 random MTM designs. 40000 of these were used as the input training data, while 5000 each were reserved for testing and validation, respectively. To obtain the corresponding labels, we analytically calculated the reflection spectra using the transfer matrix method (TMM; described further in the following section). A hypothetical example of a reflected spectrum considering an incident amplitude of unity for all frequencies is shown in Figure 2c where the solid red circles (data points) correspond to the output nodes, namely $\hat{y}_j$, from Figure 2b. We ran the training loop for 30 epochs with a batch size of 32 samples, learning rate of 0.001, ReLU activation function and Adam optimiser (beta$_1$ = 0.9, beta$_2$ = 0.999). With these parameters, the model took approximately 45 seconds to train on GPU (Nvidia GeForce RTX$^{TM}$ 3070) and converged to a MSE loss of around $\sim 5 \times 10^{-4}$ on the test set. A learning curve plot containing the loss and validation loss data for this training can be seen in Figure 3a. In this way, we obtained a fully connected NN for efficiently iterating through many design candidates and accurately predict the corresponding amplitude of the reflection spectra.

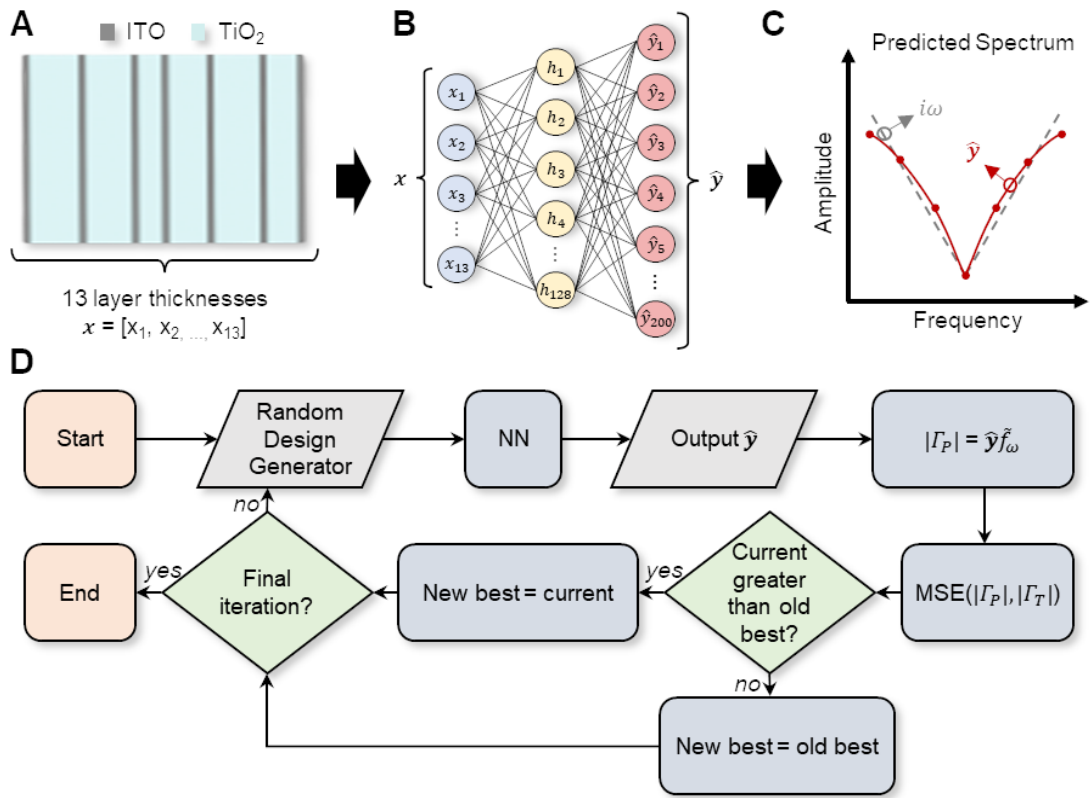

**Fig. 2 | Design process by NN and iterative algorithm.** (a) schematic representation of a multilayer MTM with thicknesses $x_i$. (b) fully connected NN diagram; the blue input nodes correspond to the MTM layer thicknesses while the red output nodes represent the predicted reflection spectrum. (c) a hypothetical predicted reflection spectrum; note that the data points relate directly to the output nodes in (b). (d) NN-based iterative algorithm where $|\Gamma_P|$ is the result of the product between the NN-predicted spectrum and the magnitude of the incident spectrum; $|\Gamma_T|$ is the target (theoretical) derivative. Note that here the product $\hat{y}\tilde{f}_\omega$ denotes the element-wise multiplication of two arrays which are discretised for the frequencies within our spectral range of interest.

*Transfer matrix method for data generation*

Given that the dimensions of the materials involved in our multilayer MTM only vary along the propagation axis, each design in the input training set is stored as a 13-element array of randomly generated layer thicknesses. The corresponding reflection spectra (training labels) are calculated using the TMM for stratified media [68]–[70]. In this approach, each layer within the MTM is modelled via a characteristic matrix, as follows:

$$\boldsymbol{M}_{TM}^{(l)} = \begin{pmatrix} \cos(d_l \beta_l) & \sin(d_l \beta_l) n_l^2 / \beta_l \\ -\sin(d_l \beta_l) \beta_l / n_l^2 & \sin(d_l \beta_l) \end{pmatrix} \qquad (2)$$

where $l = 1, 2, 3 \ldots 13$ is the layer index, $d_l$ is the thickness, $n_l$ is the refractive index and $\beta_l = k_0 n_l$ the propagation constant with $k_0 = \frac{2\pi}{\lambda_0} = \frac{\omega}{c}$ the free space wavenumber and $c$ as the velocity of light in vacuum. The total transfer matrix can then be constructed by taking the product of the characteristic matrices for each of the individual layers,

$$\boldsymbol{M} = \prod_{q=0}^{N-1} \boldsymbol{M}^{(N-q)} \qquad (3)$$

where $N$ is the number of layers (13 in our case). Finally, we take the inverse of $\boldsymbol{M}$ (total transfer matrix, left side from Eq. (3)) and use an appropriate Fresnel equation to find the reflection coefficients in the transverse magnetic (TM) polarisation,

$$r_{TM} = -\frac{M_{11}^{-1} \beta_i - M_{22}^{-1} \beta_t + i[M_{21}^{-1} + M_{12}^{-1} \beta_i \beta_t]}{M_{11}^{-1} \beta_i + M_{22}^{-1} \beta_t - i[M_{21}^{-1} - M_{12}^{-1} \beta_i \beta_t]} \qquad (4)$$

where the subscripts $i$ and $t$ refer to the media at the front and back of the multilayer, respectively (both being air in our case). With this setup, for instance, an arbitrary design array with thicknesses $\boldsymbol{x} = [x_1, x_2, \ldots, x_{13}]$ corresponds to a reflected spectrum $\boldsymbol{y} = [r_{TM,\omega_1}, r_{TM,\omega_2}, \ldots, r_{TM,\omega_{200}}]$ where each element represents the spectrum at a given frequency point as calculated by Eq. (4).

*NN-based algorithm for MTM prediction*

The derivative of a temporal function is defined in the Fourier domain as a product with the frequency variable and imaginary unit [71],

$$\mathcal{F}\left[\frac{df(t)}{dt}\right] \equiv i\omega \tilde{f}(\omega) \qquad (5)$$

Additionally, as the TMM (used here to calculate our training labels) assumes the frequency content of the incident signal to be unity for all frequencies, we can consider the output of our trained NN as the transfer function of the multilayer MTM (which we require to be $i\omega$ in order to satisfy Eq. (5)). Using this knowledge, we paired our trained NN with an iterative algorithm to test a number of candidate designs (see Figure 2d for an overview of this process). In the first step, we randomly generate a design candidate conforming to the specifications provided in section *Materials and design parameters* above. We then multiply the NN output for this design by the frequency content of the incident signal to obtain a prediction for the magnitude of the associated reflection spectrum $|\Gamma_P|$. After testing 10000 candidate designs (beyond this negligible error reduction was observed), the proposed algorithm returns the design with the lowest MSE to the frequency content for the desired derivative (here called theoretical reflected spectrum) $|\Gamma_T|$).

## Results

### *Temporal derivative of an incident modulated Gaussian signal*

With our model now trained for making efficient and accurate predictions of reflection spectra, we can set about the task of obtaining a MTM design for calculating the temporal derivative of an incident modulated Gaussian signal (with spectrum in the spectral range of ∼1428 nm and ∼1695 nm (i.e., 177 THz – 210 THz) and a central wavelength of 1550 nm). As detailed above, we iterated through 10000 design candidates of which the design with the 13-layered thickness of (in nm) [43, 208, 52, 221, 48, 247, 58, 212, 45, 290, 43, 266, 49] resulted in the lowest MSE (with a value of ∼$8 \times 10^{-4}$) between the frequency content of the theoretical temporal derivative $|\Gamma_T|$ and the product of the predicted reflection spectrum with the incident Gaussian spectrum $|\Gamma_P|$. As an additional step, we calculated the R2 score in order to evaluate the goodness of fit between $|\Gamma_T|$ and $|\Gamma_P|$ (with a value of R2 equals to 100% meaning a perfect fit between the results). This was found to be 92% in this case, demonstrating a good agreement between the two spectra. A scaled schematic representation of this design (labelled as Design A) is shown in Figure 3b. In this design, the odd and even values in the vector above (from left to right) represent the ITO and $TiO_2$ media, respectively. It is important to note that, a small frequency shift was observed between the ideal spectrum of the temporal derivative for the incident Gaussian with its minimum appearing at a wavelength of 1551.3 nm (red-shifted from the ideal modulated wavelength of 1550 nm). Similarly, the results obtained from numerical simulations (CST Studio Suite®) were also red-shifted with a central wavelength of 1552.2 nm. To overcome this, an optimization of the layer thicknesses was carried out using the previously mentioned dimensions as starting point. After this optimization (not shown), it was found that the original thickness dimensions only needed to be scaled by a factor of 0.998 for the numerical results and 0.9988 for the analytic results (with a final MSE and R2 values of $\sim 4 \times 10^{-4}$ and 95.7%), demonstrating that the deviation was a small percentage from the original dimensions. With this setup, the numerical results of the electric field distribution, as a function of space and time, are shown in Figure 3c. Here the impinging temporally modulated Gaussian signal is applied at a distance $z = 100$ $\mu$m while the multilayer MTM with Design A is placed at $z = 0$ $\mu$m.

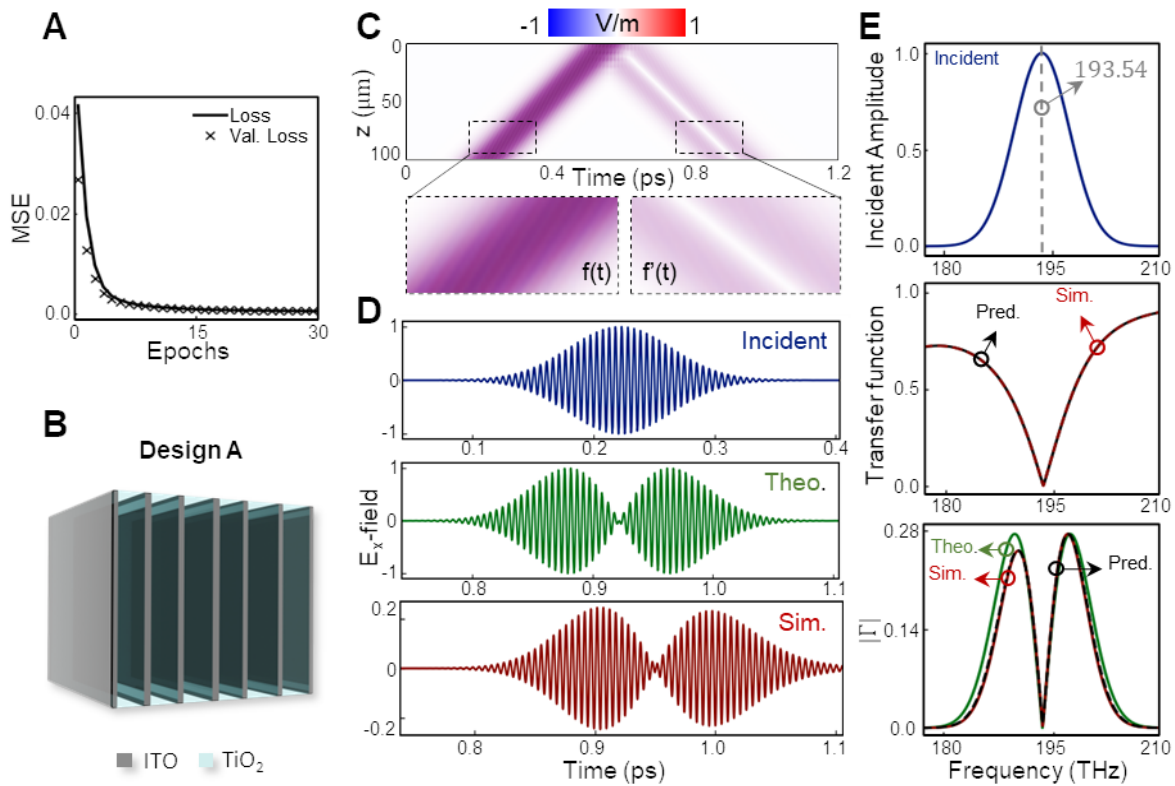

**Fig 3 | Results of the NN prediction for the Gaussian derivative.** (a) learning curve for the NN training. (b) a scale schematic of the obtained MTM design for the Gaussian derivative case (Design A). (c) full space-time plot of the incident and reflected temporal signal. The multilayer MTM is positioned at $z = 0$ $\mu$m. (d) from top to bottom, the *x*-component of the electric field for the incident Gaussian signal (blue), the theoretical derivative (green) and simulation results (red) of the reflected temporal signal for Design A. (e) top: the frequency content of the incident Gaussian signal (blue); middle: predicted (black, dashed) and simulated (red) transfer function for Design A; bottom: frequency content of the theoretical derivative (green) and simulation results (red) calculated from panel (d) along with the predicted reflection spectrum (black, dashed) obtained as the product of the incident spectrum (top) with the transfer function for Design A.

As observed, the modulated Gaussian pulse travels in free space and, after interacting with the MTM, a reflected signal is obtained which represents the temporal derivative of the incident temporally modulated Gaussian pulse. To better observe these results, we extracted the electric field distribution from Figure 3c at $z = 100$ $\mu$m (i.e., away from the multilayer MTM) and plotted the results for times between 0 ps to 0.4 ps for the incident Gaussian signal (top panel from Figure 3d, blue plot) and from 0.7 ps to 1.1 ps for the reflected signal (bottom panel from Figure 3d, red plot). The reflected temporal signal qualitatively resembles the theoretical/target time derivative (green plot from the same Figure 3d). To further analyse these results, the spectral content of the temporal signals is shown in Figure 3e. In the top panel of Figure 3e, the frequency content of the incident temporal signal is plotted, showing how its central frequency corresponds to that of a telecom wavelength of 1550 nm, as expected. For completeness, we provide in the middle panel of Figure 3e the transfer function results for Design A as predicted by our NN (black, dashed line) and numerically calculated using CST Studio Suite® (red solid line). As can be seen, an excellent agreement is achieved between these results. Moreover, note how the reflection coefficient is approximately linear close to the central frequency with a minimum at the design wavelength of 1550 nm, in good agreement with the desired transfer function $i\omega$ such as in Figure 2c. Finally, the bottom panel of Figure 3e shows the spectral response of the theoretical derivative (green, calculated from the middle panel of Figure 3d) along with the predicted reflection spectrum (black dashed line, obtained by multiplying the incident spectrum by the transfer function for Design A) and simulation results using CST Studio Suite® (red plot, extracted from the temporal signal shown in the bottom panel of Figure 3d). As observed, an excellent agreement is obtained between all the results, demonstrating how Design A is able to calculate the derivative of a temporally modulated Gaussian signal working in reflection mode. For completeness a study of the performance of this design and the bandwidth of the incident Gaussian signal can be found in the supplementary materials document.

*Temporal derivative of arbitrary functions*

To further exploit our NN-based design methodology and to evaluate more complex incident temporal signals, we implemented our algorithm again, but this time considering the profile of the incident signal to represent an arbitrary function. Note that here we opted to obtain a second MTM design in order to show that multiple distinct designs could be obtained using our approach. However, as shown in the supplementary materials document, Design A does also correctly produce the temporal derivative of an arbitrary incident signal. The newly obtained design, Design B in Figure 4a, has the following dimensions (in nm) [66, 246, 45, 200, 50, 249, 41, 261, 66, 230, 66, 208, 52] where the odd and even values (from left to right) represent the ITO and $TiO_2$ media, respectively. As with the previous case (Design A), a further optimization of the layer thicknesses was carried out. Based on this, it was found that a small scaling factor of 1.004 was enough for both numerical simulations and analytical (predicted) values to match the minimum of the spectrum corresponding to the theoretical (ideal) temporal derivative (i.e., 1550 nm as explained in the section about Design A above). With this configuration, the predicted and numerical results of the transfer function (reflection coefficient considering all input frequencies to have an amplitude of 1) are plotted in the right panel of Figure 4a, showing an excellent agreement between the results.

For this Design B, the incident temporal signal with an arbitrary envelope was based on the superposition of two modulated Gaussian functions with different relative values for amplitude (0.75 and 1) and standard deviation (80 fm and 60 fm, respectively). This arbitrary temporal signal (from now called arbitrary signal 1) can be seen in the top panel of Figure 4b (blue curve). With this setup, the theoretical and full-wave simulation results of the reflected $E_x$ as a function of time are shown in the middle and bottom panels of Figure 4b (green and red lines, respectively). As observed, there is an excellent agreement between the results, demonstrating how Design B is able to calculate the temporal derivative of the incident arbitrary signal 1. For completeness, the corresponding frequency content (from which the NN bases its predictions) for the incident signal is shown in the left panel of Figure 4c (blue line), while the right panel shows the frequency content of the theoretical derivative $|\Gamma_T|$ (green) and simulation derivative (red) calculated from the temporal signals in Figure 4b along with the predicted derivative $|\Gamma_P|$ (black, dashed) obtained as the product of the incident spectrum (on the left) with the transfer function for Design B. As can be seen, a good agreement is obtained between all the results – achieving MSE and R2 score values between $|\Gamma_T|$ and $|\Gamma_P|$ of $\sim 4 \times 10^{-5}$ and 98.6%, respectively – demonstrating how our NN-based design of multilayer MTMs can be used to calculate the temporal derivatives of arbitrary broadband sinusoidally modulated signals.

*Temporal derivative of a skyline*

As a final example and motivated by a desire to demonstrate the versatility of our predicted designs, we reused Design B with a second arbitrary incident signal. In this case, we base the envelope of the incident signal on the skyline of one of Newcastle University's most prominent structures, the Armstrong building (seen in Figure 4d). As observed, a blue line has been superimposed to this picture corresponding to the profile of the building used to generate the envelope of a signal in the time domain. Based on this profile, the incident temporal signal is plotted at the top of Figure 4e (blue plot, from now on called arbitrary signal 2) along with the theoretical temporal derivative (middle, green) and full-wave simulation results of the reflected $E_x$ as a function of time (bottom, red). As with the previous case, the frequency content of the arbitrary signal 2 is presented in Figure 4f (left), while Figure 4f (right) shows the frequency content of the theoretical and simulated derivatives calculated from Figure 4e (green and red, respectively) along with the predicted derivative (black, dashed) obtained as the product of the incident spectrum (left) with the transfer function for Design B. As observed, a good agreement is achieved between all the results for frequencies close to the modulation frequency (193.54 THz, 1550 nm wavelength) with some deviations at higher and lower frequencies. This is an expected performance due to the fact that the transfer function of the multilayer MTM is not completely linear for all the frequencies in the spectral range of interest (as observed in Figure 4a). However, in this example, a MSE between the spectral content of the analytical (predicted) and theoretical (ideal) derivatives of $\sim 6 \times 10^{-4}$ was obtained, demonstrating the ability to implement Design B for a different incident modulated temporal signal.

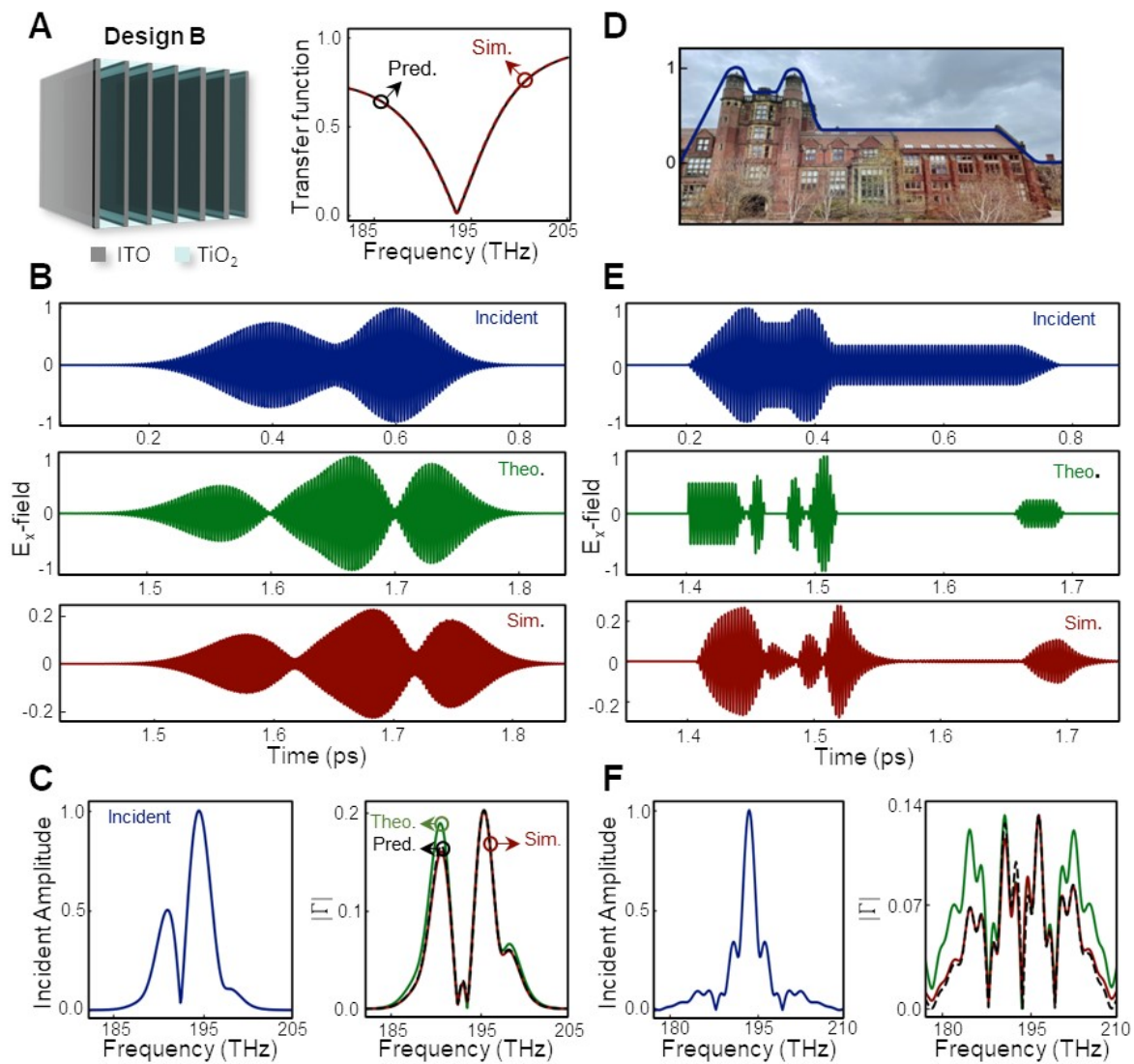

**Fig 4 | Results of the NN prediction for the arbitrary function.** (a) left: schematic diagram of Design B; right: corresponding predicted (black, dashed) and simulated (red) transfer functions for Design B. (b) top: incident arbitrary temporal signal 1; middle: theoretical temporal derivative of arbitrary signal 1; bottom: simulation results for arbitrary signal 1 based on Design B. (c) left: frequency content corresponding to incident arbitrary signal 1; right: frequency content of the theoretical derivative (green) and simulation derivative (red) calculated from panel (b) along with the predicted derivative (black, dashed) obtained as the product of the incident spectrum (on the left) with the transfer function for Design B. (d) Newcastle University's Armstrong building, the blue superimposed line corresponds to the envelope used for arbitrary signal 2. (e) top: incident arbitrary signal 2; middle: theoretical temporal derivative of arbitrary signal 2; bottom: simulation results for arbitrary signal 2 based on Design B. (f) left: frequency content corresponding to incident arbitrary signal 2; right: frequency content of the theoretical derivative (green) and simulation derivative (red) calculated from panel (e) along with the predicted derivative (black, dashed) obtained as the product of the incident spectrum (left) with the transfer function for Design B.

## Methods

*Numerical simulations*

The numerical simulations from Figures 3 and 4 were carried out using the time domain solver of the commercial software CST Studio Suite®. Waveguide ports were used at the front and back of the multilayer MTM (input/output ports). The incident signals (modulated Gaussians and arbitrary signals) where imported from external files and the incident waveguide port was used to excite the multilayer MTMs with the electric field polarized along the *x*-axis ($E_x$). Electric and magnetic boundary conditions were applied to the left/right and top/bottom boundaries, respectively. To reduce simulation time, electric and magnetic symmetries were introduced on the *yz* and *xz* planes, respectively. Finally, an extremely refined mesh was implemented with a minimum and maximum mesh size of ~$0.0016\lambda_0$ and ~$0.011\lambda_0$ with $\lambda_0 = 1550$ nm.

## Discussion

Inspired by the importance of performing mathematical operations on temporally modulated electromagnetic signals, we have made use of a NN-based iterative algorithm to design multilayer MTMs capable of taking the first temporal derivative of the envelope of an impinging electromagnetic signal modulated at telecom wavelengths (i.e. 1550 nm). Using this methodology, we have obtained a suitable MTM design, Design A, where the envelope of the reflected temporal signal correctly represents the well-known derivative of an incident temporal Gaussian pulse. This methodology was also implemented to design a second multilayer MTM, Design B, to calculate the temporal derivative of two different arbitrary temporal signals. It was shown how, after the incident signals interact with the designed MTM, the reflected signals indeed resemble their corresponding theoretical derivatives in the time domain. In all of the cases shown here, our NN-based algorithm was able to complete its search of the design space for the layer thicknesses of the multilayer MTM in just a few seconds, achieving a low MSE to the order of $10^{-4}$ or below when comparing the spectrum of the theoretical temporal derivative to the predicted reflection spectrum obtained by our NN approach. To demonstrate the versatility of the designed MTMs using our NN-based iterative algorithm, different MTM designs were tested using different modulated temporal signals, demonstrating that the reflected signals were in agreement to their corresponding theoretical derivative in the time domain. These results indicated that the transfer function of the designed MTM is consistent with the general differential operator $i\omega$ and not biased to the choice of incident signal. The methodology presented here could find important applications in designing MTMs for high-speed signal processing, temporal differentiation and wave-based equation solving, opening further opportunities to improve and accelerate the design of MTMs for computing applications.

## Conflicts of interests

The authors declare no conflict of interests.

## Acknowledgements

V.P-P. and A. Y. would like to thank the support of the Leverhulme Trust under the Leverhulme Trust Research Project Grant scheme (RPG-2020-316). V.P.-P. also acknowledges support from Newcastle University (Newcastle University Research Fellowship). V.P.-P., T. K. and A.Y. would like to thank the support from the Engineering and Physical Sciences Research Council (EPSRC) under the scheme EPSRC DTP PhD scheme (EP/T517914/1).

## Data Access Statement

All data that supports the findings of this manuscript is available in the main text. Specific data is also available from the corresponding author upon reasonable request.